\documentclass[aps,apl,twocolumn,superscriptaddress]{revtex4}

\usepackage{graphicx}
\usepackage{epstopdf}
\usepackage{comment}
\usepackage{amsmath,amssymb}
\usepackage{bm}
\newcommand{\ctext}[1]{\raise0.2ex\hbox{\textcircled{\scriptsize{#1}}}}

\begin{document}


\title{Evanescently coupled topological ring-waveguide systems for chip-scale ultrahigh frequency phononic circuits}

\author{Daiki Hatanaka}
\email{daiki.hatanaka@ntt.com}
\affiliation{NTT Basic Research Laboratories, NTT Corporation, Atsugi-shi, Kanagawa 243-0198, Japan}

\author{Hiroaki Takeshita}
\affiliation{Department of Electrical and Electronic Engineering, Okayama University, Okayama 700-8530, Japan}

\author{Motoki Kataoka}
\affiliation{Department of Electrical and Electronic Engineering, Okayama University, Okayama 700-8530, Japan}


\author{Hajime Okamoto}
\affiliation{NTT Basic Research Laboratories, NTT Corporation, Atsugi-shi, Kanagawa 243-0198, Japan}

\author{Kenji Tsuruta}
\affiliation{Department of Electrical and Electronic Engineering, Okayama University, Okayama 700-8530, Japan}

\author{Hiroshi Yamaguchi}
\affiliation{NTT Basic Research Laboratories, NTT Corporation, Atsugi-shi, Kanagawa 243-0198, Japan}

\begin{abstract}
	Topological phononics enabling backscattering-immune transport is expected to improve the performance of electromechanical systems for classical and quantum information technologies. Nonetheless, most of the previous demonstrations utilized macroscale and low-frequency structures and thus offered little experimental insight into ultrahigh frequency phonon transport, especially in chip-scale circuits. Here, we report microwave phonon transmissions in a microscopic topological ring-waveguide coupled system, which is an important building block for wave-based signal processing. The elastic waves in the topological waveguide evanescently couple to the ring resonator, while maintaining the valley pseudospin polarization. The resultant waves are robust to backscattering even in the tiny hexagonal ring, generating a resonant phonon circulation. Furthermore, the evanescently coupled structure allows for a critical coupling, where valley-dependent ring-waveguide interference enables blocking of the topological edge transmission. Our demonstrations reveal the capability of using topological phenomena to manipulate ultrahigh frequency elastic waves in intricate phononic circuits for classical and quantum signal-processing applications.
\end{abstract}
\maketitle
\begin{figure*}[t]
	\begin{center}
		\vspace{-0.cm}\hspace{-0.0cm}
		\includegraphics[scale=0.9]{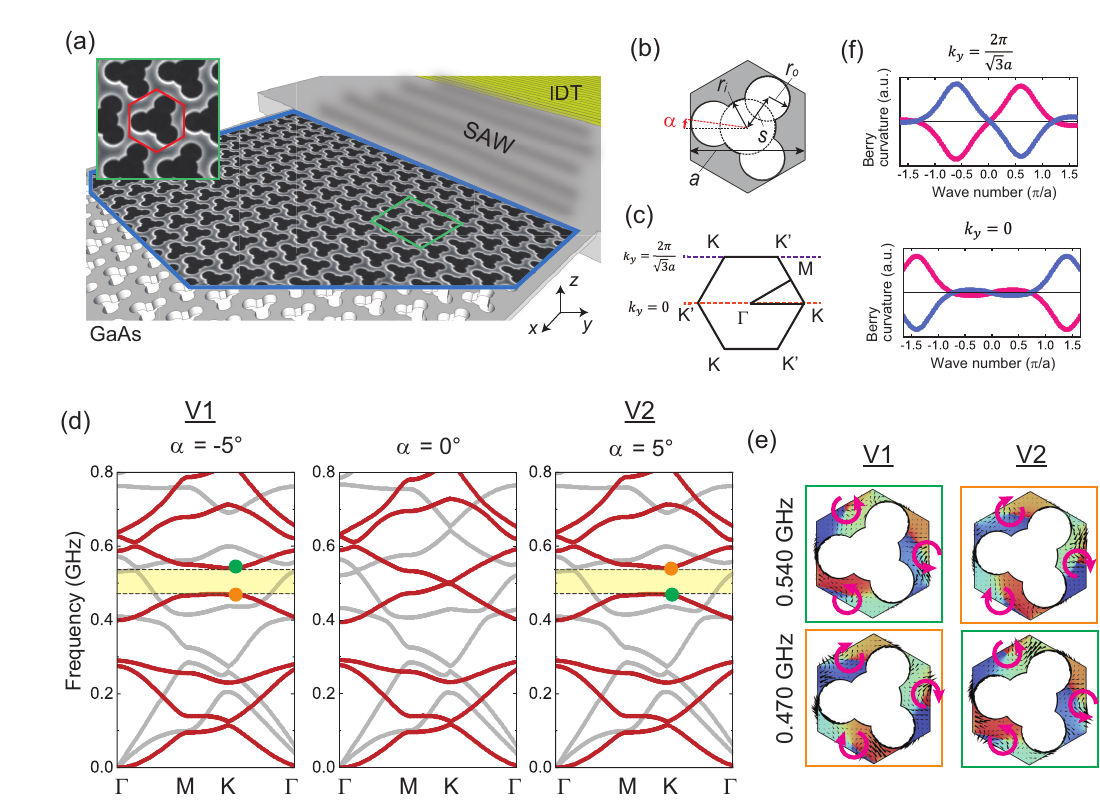}
		\vspace{0.cm}
		\caption{\textbf{(a)} Schematic diagram of a valley topological system fabricated on a GaAs suspended membrane. Surface acoustic waves (SAW) at 0.5 GHz are piezoelectrically excited by the inter-digit transducer (IDT: colored yellow) and are injected into the system (the blue bordered area). A zoomed scanning electron microscopy (SEM) image of the green bordered area of the topological phononic crystals (PnCs) is shown in the inset: unit cell contains three-fold symmetric air holes ($C_3$), as denoted by the red solid line. \textbf{(b)} and \textbf{(c)} Illustration of the unit cell and its reciprocal structure, with $a=$ 3.65 $\mu$m, $s =$ 1.10 $\mu$m, $r_i =$ 0.90 $\mu$m, and $r_o =$ 0.72 $\mu$m. $\alpha$ is the rotation angle at which inversion symmetry breaks and that governs the valley topological degree-of-freedom. \textbf{(d)} Dispersion relation of the valley Hall (VH) PnC with $\alpha =$ -5$^\circ$ (left), 0$^\circ$ (middle) and 5$^\circ$ (right). The red and gray solid lines denote out-of-plane and in-plane vibrations, respectively. The bandgap for the out-of-plane mode (asymmetric Lamb mode) emerges around 0.5 GHz (yellow) when rotating the center hole to $\alpha = \pm 5^\circ$. \textbf{(e)} Spatial distribution of out-of-plane vibration phase with frequency of 0.470 GHz and 0.540 GHz at the K point in PnCs V1 and V2, denoted by the orange and green solid circles in (d). Elastic energy flux (red arrows) circulates clockwise and counterclockwise in V1 and V2. \textbf{(f)} Numerically calculated Berry curvature of the upper band as a function of wavenumber along K-K' (top panel, $k_y = 2\pi/(\sqrt(3)a)$) and K'-$\Gamma$-K (bottom, $k_y = 0$). The red and blue solid lines are results obtained in V1 and V2, respectively.}
		\label{fig 1}
		\vspace{-0cm}
	\end{center}
\end{figure*}
\begin{figure*}[t]
	\begin{center}
		\vspace{-0.0cm}\hspace{-0.0cm}
		\includegraphics[scale=0.95]{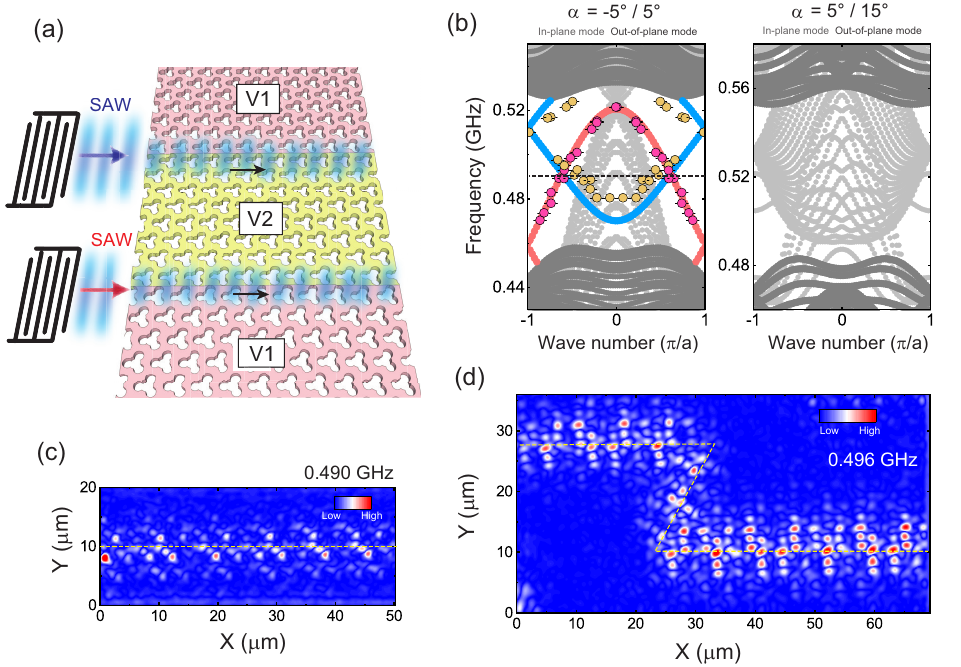}
		\vspace{0.cm}
		\caption{\textbf{(a)} Schematic illustration of valley topological edge waveguides that consist of interfaces between PnCs with opposite VH phases, V1 (red) and V2 (yellow). These boundaries are driven by injecting elastic waves from an IDT and the resultant vibrations are measured by an optical interferometer at room temperature and in a moderate vacuum \textbf{(b)} Dispersion relation of elastic waves propagating in the VH edges between $\alpha = \pm 5^\circ$ (left) and $\alpha=5, 15^\circ$ (right). Left: gapless edge modes across the bulk bandgap (dark gray) appear at the V1/V2 (blue) and V2/V1 (red) interfaces. In-plane vibrations (light gray) simultaneously exist within the bandgap, but they cannot be detected due to the limitations of the optical measurement scheme. Experimental dispersion relations are plotted as solid yellow and red circles. Right: the gapless edge modes disappear at the boundaries made by phononic lattices with the same sign of $C_v$, i.e. $\alpha = 5^\circ$ and $15^\circ$. \textbf{(c)} and \textbf{(d)} Spatial profile of elastic waves propagating through a straight (V1/V2) and a Z-shaped (V2/V1) valley topological edge, respectively. Injected elastic waves are confined around the inverted VH boundary and smoothly propagate through the two sharp bends without significant wave localization.}
		\label{fig 2}
		\vspace{-0cm}
	\end{center}
\end{figure*}
\section{Introduction}
\hspace*{0.cm}Topological insulators represent phases of matter, where two topologically distinguished bulk phases guarantee the existence of lower-dimensional edge states at the boundary between them \cite{thouless1982quantized, hasan2010colloquium, qi2011topological}. This phenomenon, known as the bulk-boundary correspondence, allows robust and backscattering-immune carrier transport that is unaffected by structural defects and fabrication imperfections. This exotic topological property was first studied in electronic systems and later in a variety of platforms for studying electromagnetic waves \cite{wang2009observation, chen2014experimental, cheng2016robust}, light waves \cite{rechtsman2013photonic, hafezi2013imaging, he2019silicon, shalaev2019robust, rechtsman2023reciprocal}, cold atoms \cite{jotzu2014experimental, aidelsburger2015measuring}, plasmon \cite{gao2016probing, wu2017direct}, acoustic \cite{yang2015topological, xiao2015geometric, he2016acoustic, he2020acoustic} and elastic waves \cite{lu2017observation, cha2018experimental, yu2018elastic, painter_topo_pnc, yan2018chip, yu2021critical, zhang2022gigahertz, nii2023imaging}. The findings of these studies have provided a foundation for advanced manipulations of these classical waves.\\
\hspace*{0.4cm}Of particular interest is phononic systems, which are well-suited for rigorous structural design and fabrication due to the millimeter-scale wavelengths of sound and ultrasound waves. Mechanical fabrication techniques, such as metalworking and additive manufacturing, have been used to construct intricate topological acoustic structures in two and three dimensions (e.g. ref. \cite{yu2018elastic, he2020acoustic}) and have enabled observation of topologically protected phonon transport in accordance with theoretical predictions. These topological characteristics are highly desired as they would enable spatially fine and precise manipulations of gigahertz elastic waves. This has been a significant challenge in conventional, topologically-\textit{trivial} systems because such ultrahigh-frequency elastic waves, with nano- and micrometer-scale wavelengths, are particularly susceptible to backscattering due to structural disorder, thereby resulting in the waves being localized accidentally in real and frequency space. The topological approaches in microwave phononics are expected to overcome this difficulty and substantially enhance both the controllability and adaptability to integrated circuits of ultrahigh-frequency phonons.\\
\hspace*{0.4cm}In spite of the promise, experimental demonstrations on ultrahigh frequency and microscopic systems have been limited to a few reports \cite{zhang2022gigahertz, nii2023imaging}, in which topological phonon transport was only investigated in simple valley Hall (VH) edge waveguides because of complexity of their fabrication and measurement setups. However, the benefits of topological phononics are not restricted to these sorts of waveguide methods and can be had by integrating waveguides with other components, such as resonators, splitters, and modulators. Ring resonators and their corresponding waveguide coupling systems are especially noteworthy because their unique characteristics arising from critical coupling phenomena can be used to control valley phonon transport by filtering, switching, and multiplexing \cite{yu2021critical} as well as phononic analogue of topological lasing \cite{peano2016topological, harari2018topological, bandres2018topological}. In addition, valley pseudospins and orbital angular moments of ultrahigh frequency phonons in the ring can interact with magnons and photons \cite{thingstad2019chiral, fu2019phononic}, which may be useful in making chiral phonon-based hybrid systems. The phonon transport property of these ideal systems is free of significant backscattering at corners and defects \cite{lu2017observation, cha2018experimental, yu2018elastic, painter_topo_pnc, yan2018chip, yu2021critical, zhang2022gigahertz, nii2023imaging}; this will enable construction of functional systems with compact wavelength-scale structures that have both high performance and design flexibility. Thus, exploiting the concepts of topology is promising for enhancing the capability and versatility of integrated phononics technology at microwave frequencies.\\
\hspace*{0.4cm}Here, we developed VH topological ring-waveguide integrated systems whose area confining the elastic waves is eight orders of magnitude smaller than that of previous ultrasonic topological structures \cite{yu2021critical}. Spatial mapping of the vibration amplitude revealed that valley polarized elastic waves at 0.5 GHz in the waveguide evanescently couple to the ring resonator and selectively excite valley-dependent ring resonances. Even in such a ultrahigh frequency regime, wave propagation is surprisingly smooth without being disturbed by the ring corners. In contrast, similarly fabricated but topologically trivial systems show strong wave localization at the corners. Furthermore, the VH ring-waveguide systems showed critical coupling, where the incoming waves interfere with the circulating waves with the same valley polarization, resulting in the transmission of the topological waves being blocked. Our findings show the capability of topological technology for manipulating microwave phonons in on-chip integrated circuits in classical and quantum information processing applications \cite{zhang2022gigahertz}.\\
\section{Fundamental properties of valley Hall topological phononic crystals}
\subsection{Valley Hall topological phononic crystal}
\hspace*{0.4cm}The VH phononic crystals (PnC) were formed in an under-etched GaAs membrane and consisted of a triangular lattice of hexagonal unit cells with three-fold symmetric holes, as shown in Fig. 1(a) and 1(b). This engineered structure is known to have a graphene-like dispersion relation for phonons \cite{yu2016surface}. Figures 1(c) and 1(d) show the reciprocal unit cell and dispersion relation calculated by using the finite-element method (FEM), where phononic branches for out-of-plane elastic waves, i.e. asymmetric Lamb modes denoted by red solid lines, have Dirac degeneracy at the K (K') point and around 0.5 GHz (see the middle panel of Fig. 1(d)). This phononic graphene is a semi-metal, but it can be turned to an insulator by breaking the inversion symmetry of the unit cells \cite{yan2018chip, zhang2022gigahertz, nii2023imaging, kim2019design, ali2023reconfigurable}, whereby the interior holes are rotated around the normal axis by $\alpha = -5^{\circ}$ (V1) and $5^{\circ}$ (V2) (see the left and right panels of Fig. 1(d)). The slight modification to the cell geometry lifts the degeneracy and opens an energy gap between two phononic eigenstates holding clockwise and counterclockwise energy vortexes, as depicted in Fig. 1(e)). These PnCs sustain finite Berry curvature around the K and K' points (Fig. 1(f)), so integration around the points gives a nonzero valley Chern number $C_{v}$, whose sign is inverted between K and K'. Elastic waves experience a pseudo-magnetic field determined by the Berry curvatures as they propagate, and thus, the phonon transport is valley pseudospin-polarized and backscattering is suppressed due to conservation of time reversal symmetry. The sign of $C_{v}$ is also inverted between V1 and V2 (see Fig. 1(e)) and its topological phase transition confirms that the PnCs behave as VH topological insulators operating in an ultrahigh frequency regime.\\
\subsection{Topologically protected edge transport}
\begin{figure*}[t]
	\begin{center}
		\vspace{-0.cm}\hspace{-0.0cm}
		\includegraphics[scale=1.0]{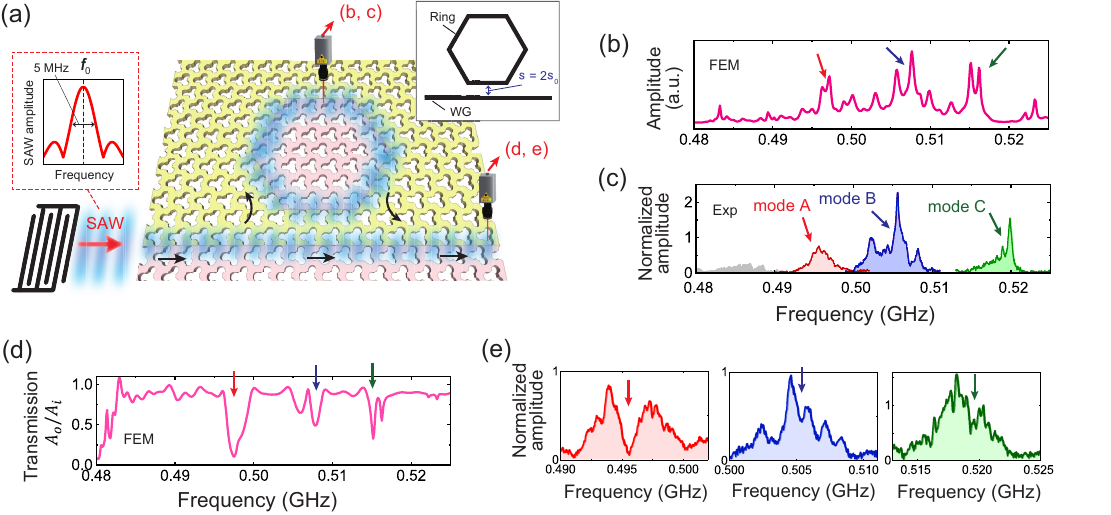}
		\vspace{0.cm}
		\caption{\textbf{(a)} Schematic diagram of valley topological hybrid system comprising a hexagonal loop ring evanescently coupled to a straight edge with a separation of two unit layers ($s/s_0=$ 2). Elastic waves are injected into the left end of the waveguide and drive the ring. Transmitted waves are optically measured at the ring and right side of the waveguide. \textbf{(b)} and \textbf{(c)} Simulated and experimental frequency responses of vibrations in the ring. Three resonant vibration peaks occur at 0.497 GHz (red arrow), 0.506 GHz (blue arrow), and 0.519 GHz (green arrow), which are reproduced in the FEM simulation. These peaks are labeled as modes A, B, and C, respectively. \textbf{(d)} Spectral response of waveguide transmission simulated by FEM. It has three dip structures at the frequencies where the ring resonances are generated (see (b)). The vibration amplitude at the output is normalized by the amplitude at the input. \textbf{(e)} Experimentally measured amplitude of the waveguide transmission around the frequencies of modes A (left), B (middle), and C (right). The excited surface acoustic waves (SAW) are spectrally limited by the finite bandwidth ($\sim$ 5 MHz) of the IDT (see the left inset of (a)), which is determined by the finger electrode period ($p=$ 5.86 $\mu$m, 5.74 $\mu$m, and 5.58 $\mu$m) and electrode array number ($N=$ 100).}
		\label{fig 3}
		\vspace{-0cm}
	\end{center}
\end{figure*}
\hspace*{0.4cm}Topologically protected guided modes emerged at the interface of the VH PnCs with inverted symmetry (opposite signs of $C_v$), V1 and V2, as shown in Fig. 2(a), where bulk-boundary correspondence guaranteed their existence. Two gapless helical edge states supporting out-of-plane vibrations appeared across the bandgap at the V1/V2 interface (blue plot in the left panel of Fig. 2(b)) and V2/V1 interface (red plot). The edge states were sustained by the nontrivial bulk topology as they would vanish if the interfaces were topologically trivial that consisted of two crystals with the same sign of $C_{v}$, e.g. $\alpha=5^\circ$ and $15^\circ$ (see the right panel of Fig. 2(b)).\\
\hspace*{0.4cm}Real-space mapping measurements were made on elastic wave transmissions piezoelectrically excited by an inter-digit transducer (IDT) at 0.490 GHz and 0.496 GHz (Fig. 2(c) and 2(d)). They showed robust topological transport, wherein vibrations confined to the boundary propagated along it without significant backscattering at the two sharply bent corners. Furthermore, the spatial mode profiles were measured at various frequencies, and they were used to experimentally plot the dispersion relation of the propagating phonons (the solid circles in the left panel of Fig. 2(b)). The experimental dispersion curves are in good agreement with the numerical ones, indicating that our semiconductor nanofabrication technology provides a good phononic platform for studying ultrahigh frequency valley topological transport.\\
\section{Topological ring resonator-waveguide coupled systems}
\subsection{Frequency response of the coupled systems}
\hspace*{0.4cm}A ring resonator-waveguide coupled structure is a key component in wave-based computation systems \cite{ferrera2010chip, liu2016fully, marpaung2019integrated}. It has been studied and used in fields of photonics and optics for a long time, and has enabled demonstrations of a range of information processing functions such as filtering, switching, multiplexing and lasing. Therefore, this realization of hybridization in a phononic system represents an important step towards constructing microwave phonon integrated circuits.\\
\begin{figure*}[t]
	\begin{center}
		\vspace{-0.cm}\hspace{-0.0cm}
		\includegraphics[scale=0.9]{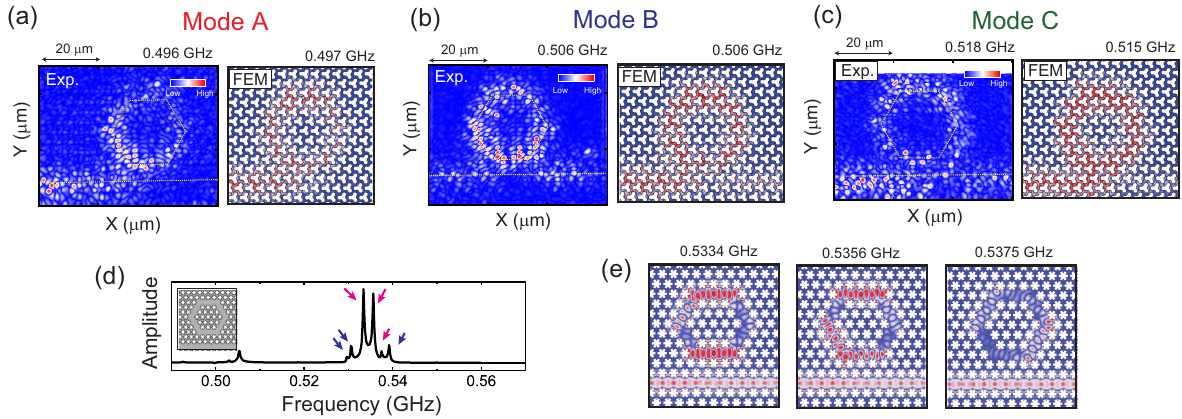}
		\vspace{0.cm}
		\caption{\textbf{(a)}-\textbf{(c)} Left panel: spatial mode profile of ring resonances experimentally measured at 0.496 GHz (mode A), 0.506 GHz (mode B), and 0.518 GHz (mode C); right panels: behavior of the ring-waveguide system as simulated using FEM at 0.497 GHz, 0.506 GHz and 0.515 GHz. \textbf{(d)} Spectral response of a topologically trivial ring formed by line-defect PnC waveguides (inset), simulated by FEM. Six resonance peaks appear (denoted by arrows) where vibrations are spectrally and spatially localized inside the ring due to strong backscattering at the corners. \textbf{(e)} Spatial mode profiles of localized resonant peaks indicated by the pink arrows in (d).}
		\label{fig 4}
		\vspace{-0cm}
	\end{center}
\end{figure*}
\hspace*{0.4cm}We formed a hexagonal ring with a radius of 2.2$\lambda$ = 12 $\mu$m (elastic wavelength $\lambda$), by enclosing the edge channel upon itself. The ring was evanescently coupled to the edge waveguide with a separation of two unit layers ($s=2s_0$ with $s_0=\sqrt{3}a/2$) as shown in Fig. 3(a). In contrast to the input port which was located on the left side of the waveguide at a distance of 65 $\mu$m from the ring, the output port was placed 530 $\mu$m at the right end from it so that waves reflected at the right end of the waveguide end would be sufficiently suppressed to avoid their interfering with waves out of the ring. Also, an IDT comprised of 100 periods of electrode pairs ($N=$ 100) was formed on the bulky substrate area 620 $\mu$m from the input port. This configuration allowed us to excite strong elastic waves and sensitively measure them by using a time gate technique and an optical interferometer, in which only delayed elastic signals are detected by filtering out fast-traveling electrical cross-talk signals. In this way, our VH PnC platform has been optimized to making measurements on intricate phononic circuits.\\
\hspace*{0.4cm}The ultrahigh frequency elastic waves incident on the left input port excited valley pseudospin-locked elastic waves in the edge waveguide and these valley pseudospin-locked waves propagated into the ring to cause ring resonant oscillations. Figures 3(b) and 3(c) show numerically calculated and optically measured spectral responses of the vibrations in the VH topological ring between 0.480 and 0.523 GHz. Both the responses show that the ring supports three elastic resonances. Hereafter, these modes appearing around 0.496 GHz, 0.506 GHz and 0.518 GHz are called modes A, B and C.\\
\hspace*{0.4cm}The spectral responses of the waveguide transmission was numerically investigated in the same frequency range as above. The amplitudes of the transmitted vibrations in the waveguide, normalized by the amplitude of the wave incident at the input, are plotted as function of frequency in Fig. 3(d). The dips in this plot correspond to those where the ring resonances of modes A, B, and C occur (see Fig. 3(b)), so the transmission suppression is a consequence of the energy consumed by driving the ring. The experimentally measured spectral responses of the edge transmission around the frequencies of modes A, B, and C are shown in Fig. 3(e). Again, dip structures appear at these mode frequencies within a finite transmission bandwidth of nearly $f_0 / N =$ 5 MHz (this value is limited by the IDT's geometry, i.e., its electrode period ($p$) and period number ($N$)). In the case of mode A, the amplitude of vibration in the waveguide approaches zero; this is a signature of a critical coupling between the two components. These results confirm that the ring-waveguide coupled circuit operated in accordance with the numerical model.\\    
\subsection{Real-space characterization of transmission dynamics}
\hspace*{0.4cm}To get a deeper insight into the wave propagation dynamics, we measured the spatial profile of the modes by scanning the laser spot of the optical interferometer over the coupled systems with $s=2 s_0$. As shown in the left panels of Fig. 4(a)-4(c), the elastic waves in the ring extended around its perimeter and caused valley pseudospin-momentum locked resonances. The elastic waves in mode A were unable to propagate in the waveguide due to interference from the ring resonance. Although relatively weak, a similar phenomenon was observed in mode C. Mode B showed an intense resonant circulation, which, as discussed later, was caused by suppressed inter-valley scattering. These spatial wave profiles are consistent with the numerical calculations shown in the right panels.\\
\begin{figure*}[t]
	\begin{center}
		\vspace{-0.cm}\hspace{-0.0cm}
		\includegraphics[scale=1.0]{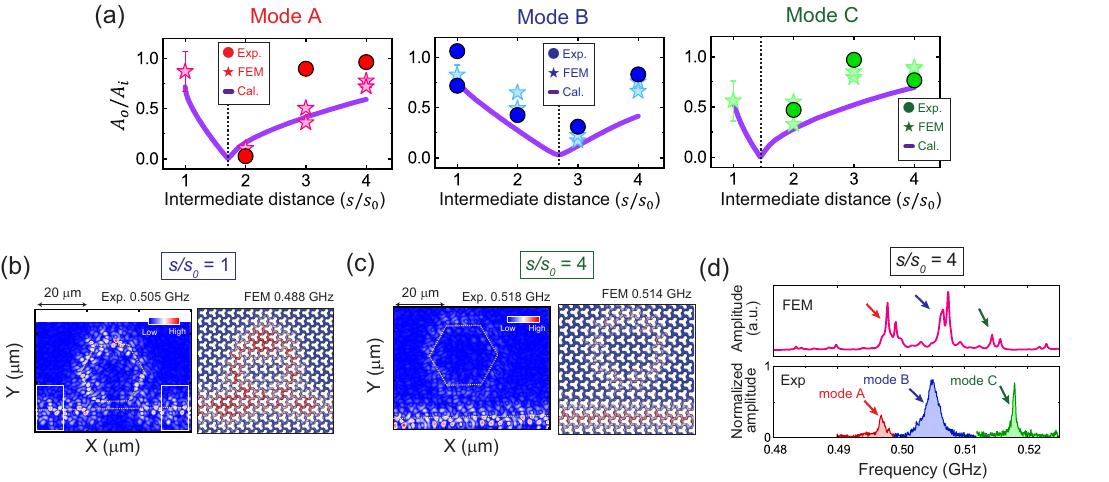}
		\vspace{0.cm}
		\caption{\textbf{(a)} Normalized amplitude ($A_o/A_i$) of the waveguide transmission as function of separation ($s/s_0$) between the ring and waveguide with $s_0=\sqrt(3)a/2$ in mode A (left), B (middle) and C (right), where $A_i$ and $A_o$ are estimated by integrating the displacement amplitudes over the input and output area (the white dotted square in (b)). The experimental and numerical simulated (FEM) results are plotted as solid circles and stars, respectively. Solid purple lines are for the analytical model based on the one of Yariv (see the Appendix). \textbf{(b)} and \textbf{(c)} Experimental (left) and numerical (right) spatial mode profiles of the coupled ring-waveguide systems in $s/s_0=$ 1 and $s/s_0=$ 4. The white dotted squares are the areas used in the estimation of $A_i$ and $A_o$. \textbf{(d)} Numerical and experimental spectral response of the ring in the coupled system with $s/s_0=$ 4.}
		\label{fig 5}
		\vspace{-0cm}
	\end{center}
\end{figure*}
\hspace*{0.4cm}We also performed numerical simulations on a topologically \textit{trivial} integrated system to investigate its effect of the nontrivial topology. This ring resonator-waveguide coupled system consisted of line-defect PnC waveguides carrying single-mode elastic waves around 0.5 GHz as shown in the inset of Fig. 4(d) \cite{hatanaka_hyperPnC, hatanaka2023phononic}. The results show that elastic waves injected from the waveguide into this ring would be strongly scattered at each corner of the hexagonal ring and become spatially and spectrally localized (see Fig. 4(d) and 4(e)). By contrast, the valley topological system avoided backscattering at the corners even through the size of the ring resonator was comparable to the wavelength and allowed less-scattered (or one-way) propagation through the ring.\\
\subsection{Ring-waveguide separation dependence of phonon transmission}
\hspace*{0.4cm}A change in the layer number ($s/s_0$) between the ring and waveguide alters the coupling strength, which allows for regulation of the edge phonon transmission as well as investigations of the valley-dependent coupled ring dynamics. We measured the normalized elastic vibration amplitude in the waveguide transmission at the frequencies of modes A, B and C while changing intermediate layer number from $s/s_0=$ 1 to 4. The results are shown in the left, middle and right panels of Fig. 5(a), where the experimental and FEM simulated results are plotted as solid circles and stars, respectively. When the ring and waveguide were close to each other ($s/s_0=1$), the incident waves could not keep going straight in the waveguide because the VH PnC (V2) no longer formed. Instead, the waves propagated into the ring in a clockwise direction and finally reached the waveguide output port while maintaining the valley pseudospin polarization rather than returning to the input port via inter-valley scattering. The results shown in the left and right panels of Fig. 5(b) also show an $\Omega$-shaped transmission profile due to pseudospin-momentum locking. This wave behavior is different from that in conventional ring-waveguide coupled systems, indicating that the topological protection functions for transport.\\  
\hspace*{0.4cm}Increasing the distance between the ring and waveguide weakened their coupling, while a VH PnC (V2) formed in the coupling region, significantly modulating the ring resonance properties. When the phonon transport was topologically protected and their valley polarization was preserved through evanescent coupling in the ring and waveguide, circulated waves in the ring destructively interfere with those in the waveguide. This results in the waveguide transmission being strongly suppressed as observed especially in mode A and $s/s_0=2$ (see Fig. 4(a) and 5(a)), where internal dissipation rate of the ring becomes equal to that of the coupling that realizes critical coupling between the two components. A further increase in separation to $s/s_0=4$ caused the ring resonator to be spectrally isolated from the waveguide (Fig. 5(d)). The weak coupling decreased the dissipation into the waveguide and increased the quality factor upto nearly $Q =$ 2,000 which is confirmed by the Lorentzian fitting to these modes in Fig. 5(c) and 5(d). On the other hand, the decoupling from the ring resulted in an increment in the waveguide transmission (see Fig. 5(a)).\\
\hspace*{0.4cm}These experimental behaviors resembled those observed in optical ring-waveguide coupled systems. Optical waves in a waveguide are transmitted through an evanescently coupled ring with their momentum preserved and form a resonant circulation, known as the whispering-galley mode (WGM), in a large-scale optical resonator. Accordingly, we decided to use an analytical approach based on the Yariv model, which is commonly used to simulate optical ring-waveguide coupled systems and to evaluate the coupling distance dependency (see the Appendix for details) \cite{yariv2002critical}. The solid lines in Fig. 5(a) indicate the normalized transmission amplitude plotted as a function of coupling distance ($s$) as solved by the analytical model. Adjusting the ring-waveguide separation realized a critical coupling at certain distances dependent on the resonant modes, which is consistent with the experimental and FEM findings. In contrast to optical systems where the diameter of ring resonator is large enough to allow the momentum preserved coupling, our topological phononic ring resonator is on the scale of a few wavelengths but supports the topological valley-conserved evanescent coupling and ring resonant circulation. This difference indicates another benefit of topological systems: minimization of the size of wave-based signal processing circuits.\\
\section{Discussion}
\hspace*{0.4cm}As shown above, our topological ring resonators have much less backscattering compared with those made from conventional PnCs. To further explore the hexagonal ring's capability of topologically protected valley polarized phonons, let us direct our attention to Fig. 5(a), in which the critical coupling configuration in mode B is realized at large separation ($s/s_0 \sim 2.7$) compared with in modes A and C ($s/s_0 \sim$ 1.7 and 1.4, respectively). The results indicate that phonon circulation loss of mode B is lower than the others, thus giving rise to well-defined and intense resonant oscillations in that mode, as shown in Fig. 3(c) and 4(b). Our analytical approach using the Yariv model also suggests that elastic wave scattering between two opposite valleys induces splitting of their spectral peaks, whose strength is weaker in mode B than in modes A and C (see the Appendix for the details). We attempted to clarify the possible origins of the inter-valley scattering with FEM and found that they could be caused by two dominant factors: spatial symmetry-breaking by the elastic stiffness of the GaAs medium and/or of the ring loop geometry. The former is attributed to the crystalline structure of GaAs (001) having four-fold rotational symmetry around the z-axis. The elastic anisotropy lifts the degeneracy of clockwise and counterclockwise resonant circulations, thus resulting in slight inter-valley scattering across all three ring resonance modes A, B and C. In the latter, the ring has six corners where the translational symmetry of the periodic structure is locally perturbed, and thereby, inter-valley scattering occurs there \cite{yu2021critical}. This process is notably manifested at certain conditions determined by the relationship between the elastic wavelength and the ring circumference, which could be satisfied with mode A and C, but not mode B (see Appendix). Consequently, we believe that the mode-dependent inter-valley scattering events results in lower dissipation in mode B than in modes A and C.\\
\hspace*{0.4cm}Finally, we should mention the effect of structural disorder induced by fabrication uncertainties on the valley topological protection. Recently, spatial transport characteristics in VH topological photonic crystals have been reported in which nanoscale structural inhomogeneities caused backscattering of light and thus, undesired spatial and spectral localization at random positions in photonic edge waveguides \cite{rechtsman2023reciprocal}. In contrast, such spatial and spectral wave localizations are relatively small in our VH PnC systems, which can be confirmed from the measured spatial mode profiles (Figs. 4(a)-(c), and 5(b)-5(c)). This indicates that the topological protection relying on time-reversal symmetry works for phonon propagation in sub-GHz and even a few GHz regimes and confirms the capability of operating chip-scale phononic circuits.\\
\section{Conclusion}
\hspace*{0.4cm}We demonstrated on-chip manipulation of ultrahigh-frequency phonon waves in a VH phononic ring-waveguide hybrid system. Valley pseudospin-locked elastic waves propagating in an edge waveguide were transmitted to a ring through evanescent coupling and excited multiple ring resonances. Their spatial mode profiles revealed that topologically protected elastic waves circulated in the tiny wavelength-scale hexagonal ring without experiencing wave localization due to strong backscattering at the corners. Moreover, the valley-dependent ring resonances created a critical coupling to the edge waveguide and enabled the transmission to be blocked through a valley preserved evanescent coupling. Our demonstrations unveiled the ability to manipulate valley-polarized phonons at ultrahigh frequencies, with which researchers can explore new fields of chiral phononics via the magnon-phonon interaction \cite{hatanaka2023phononic} and develop integrated phononics technology for classical \cite{yasuda2021mechanical} and quantum information processing technologies \cite{delsing20192019, clerk2020hybrid, dumur2021quantum}.\\
\section*{Methods}
\hspace*{0.4cm}\textbf{Fabrication.} A phononic lattice of a VH topological insulator was patterned in an epitaxially grown GaAs single crystal (100) with a thickness of 1.0 $\mu$m by standard electron beam lithography (JEOL JBX-6300) and inductively coupled plasma reactive ion etching (ULVAC NE-550). The sample was immersed in dilute hydrofluoric acid (HF 5$\%$) for 5 minutes, which allowed the perforated GaAs layer to be suspended by selective wet etching of a sacrificial layer Al$_{0.7}$Ga$_{0.3}$As with a thickness of 3.0 $\mu$m through the patterned holes. Inter-digit transducers (IDTs) were fabricated on the bulky GaAs/Al$_{0.7}$Ga$_{0.3}$As area outside of the suspended GaAs slab. The IDTs had $N=$ 100 periodically arrayed electrode finger pairs separated by a distance $p$. The spectral width for efficient piezoelectric transduction in the IDT was limited to approximately $f_{0}/N=$ 5 MHz, where $f_0$ is the center frequency determined by the surface acoustic wave (SAW) velocity ($v_{saw}$) divided by the electrode period ($f_0 = v_{saw}/p$). We prepared various IDTs with $p$ values from 5.58 $\mu$m to 5.98 $\mu$m in order to acquire broad spectra ranging from 0.480 GHz to 0.525 GHz.\\
\hspace*{0.4cm}\textbf{Measurements.} Application of an alternating voltage from a network analyzer (Keysight Technology, E5080A) or a signal generator (Anritsu, MG3740A) to the IDT generated SAWs via piezoelectric effect, which were, in turn, transformed into asymmetric Lamb waves in the non-patterned GaAs slab. The elastic waves were injected perpendicularly into the VH PnCs which allowed for an efficient coupling to the K valley phonon state at the boundary between two PnCs with opposite VH phases. The valley pseudospin-locked elastic waves were detected by an optical interferometer (Neoark, MLD-100), which converted them into electrical signals in a photo-detector. In the frequency response measurements (Fig. 3(c), 3(e) and 5(d)), these electrical signals were collected and measured in the network analyzer by using time-domain gating analysis to remove unwanted electrical cross-talk. In the real-space mapping measurements (Fig. 2(c), 2(d), 4(a)-4(c), 5(b) and 5(c)), they were frequency down-converted to 10.7 MHz and then bandpass filtered before being measured by a lock-in amplifier (Standard Research systems, SR844). All experiments were performed at room temperature and in moderate vacuum (10-100 Pa).\\
\hspace*{0.4cm}\textbf{Finite-element method.} We conducted a series of numerical simulations with COMSOL Multiphysics. The VH topological PnC was designed in an anisotropic GaAs membrane whose detailed geometric parameters are given in the caption of Fig. 1. The crystal orientations [110] and [-110] were aligned to the x- and y-axes and the density and elastic stiffness values were taken from the literature \cite{adachi_gaas}. The internal elastic dissipation of the GaAs layer was set to $Q=$ 2,000 and perfectly matched layers were set outside of the VH topological PnC area in order to suppress undesired wave reflections. In the simulations, external periodic forces were applied to the left end of the edge waveguide and the out-of-plane displacement amplitudes were calculated in the ring and on the right side of the waveguide.\\
\section*{Acknowledgments}
This work was partially supported by JSPS KAKENHI(S) Grant Number JP21H05020 and JP23H05463.\\
\section*{Author contributions}
D.H. fabricated the device and performed the measurements and the data analysis. D.H., H.T. and M.K. conducted the FEM simulations. D.H. wrote the manuscript with help of the H.Y.. All authors discussed the results during preparation of the paper.\\
%
%

%
%
\vspace{0.5cm}
\begin{appendix}
\section{Detailed device configuration}
\hspace*{0.4cm}A hexagonal ring resonator was placed at the side of a straight edge waveguide (within V2 region) at a distance of 65 $\mu$m from the left edge (input port), as shown in Fig. 6. Its output port at the right end was located far away from the ring (530 $\mu$m) in order to suppress reflected waves from it that would induce undesired vibrations (interference) in the ring. An IDT was formed on a bulky GaAs/AlGaAs substrate and located 620 $\mu$m from the input port. The long distance between the IDT and input port caused the arrival of the incident elastic waves at the ring to be delayed by 200-300 ns, so that we could investigate the responses of the ring and waveguide output by measuring their frequency spectra at A and B (see Fig. 6); here, the time-gating technique available in the network analyzer (E5080A) allowed us to filter out electrical cross-talk signals and detect only the delayed elastic ones.\\
\begin{figure*}[h]
	\begin{center}
		\vspace{-0.cm}\hspace{-0.0cm}
		\includegraphics[scale=0.8]{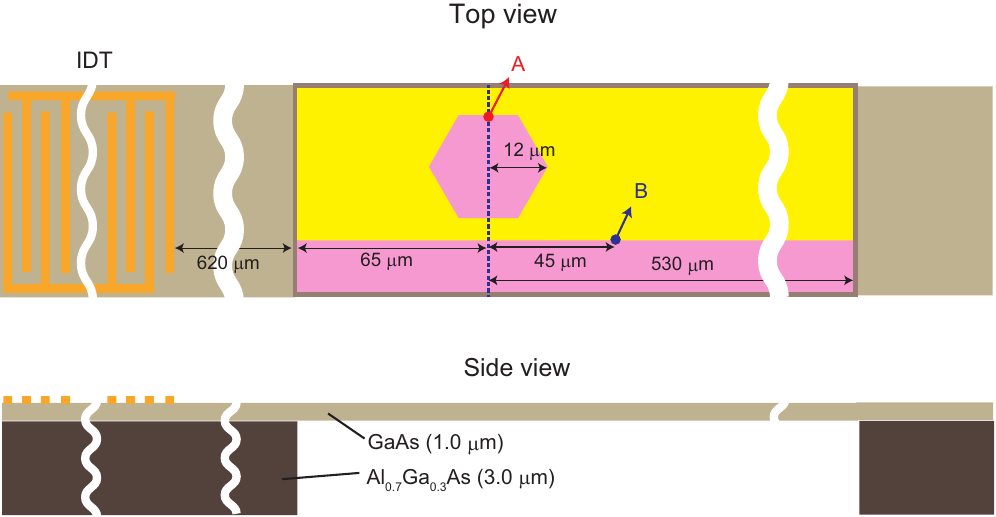}
		\vspace{0.cm}
		\caption{Schematic device configuration; top and side views are shown in the top and bottom panels, respectively. VH PnCs (V1 and V2) were under etched and suspended from the AlGaAs substrate. The spectral responses of the ring and waveguide were measured at A and B, respectively.}
		\label{fig S1}
		\vspace{-0cm}
	\end{center}
\end{figure*}
\section{Analytical solution of coupled ring-waveguide systems}
\hspace*{0.4cm}An analytical solution describing the dynamics of an optical ring-waveguide coupled system has been proposed by Yariv \cite{yariv2002critical}. We modified the Yariv's model by introducing an inter-valley scattering coefficient ($r$) that induces an interaction between the clockwise and counterclockwise circulations, as shown in Fig. 7(a). Using this model, the amplitude of vibrations in our coupled system when incident waves with amplitude $a_1$ are injected from a left port of the waveguide can be expressed\\
\\
\begin{widetext}
	\vspace{0.2cm}
	\(
	\left(
	\begin{array}{c}
	b_1 \\
	b_{cw} \\
	b_{ccw} \\
	a_{cw} \\
	a_{ccw}
	\end{array}
	\right)
	\)
	\(
	=\left(
	\begin{array}{ccccc}
		1 & -\kappa_1 & 0 & 0 & 0 \\
		0 & 1 & -ire^{(ik-\alpha)L_e} & -te^{(ik-\alpha)L_e} & 0 \\
		0 & 0 & 1 & 0 & t_1 \\
		0 & t_1 & 0 & 1 & 0 \\
		0 & 0 & -te^{(ik-\alpha)L_e} & -ire^{(ik-\alpha)L_e} & 1
	\end{array}
	\right)^{-1}
	\)
	\(
	\left(
	\begin{array}{c}
	t_1 a_1 \\
	0 \\
	0 \\
	\kappa_1 a_1 \\
	0
	\end{array}
	\right)
	\),
	\vspace{0.2cm}
\end{widetext}
where $k$, $L_e$, and $\alpha$ are the wavenumber of the elastic waves, the effective length of the ring perimeter, and the propagation loss, respectively. $a_i$ and $b_i$ ($i =$ cw, ccw) indicate the incoming and outgoing elastic waves with clockwise and counterclockwise directions in the ring as described in Fig. 7(a). These waves circulating in opposite directions interact via the scattering coefficient ($r$). The scattering and transmission coefficients in the ring ($r$ and $t$) , and  coupling and transmission coefficients in the ring-waveguide area ($\kappa_1$ and $t_1$) satisfy the following relations,\\
\begin{align}
	\left| r \right|^2 + \left| t \right|^2 &= 1, \\
	\left| \kappa_1 \right|^2 + \left| t_1 \right|^2 &= 1.
\end{align}
Upon solving these equations, we can plot the normalized transmission amplitude ($A_0/A_i = \left| b_1/a_1 \right|$) in the waveguide with intermediate layers $s/s_0=$ 2 in modes A, B, and C shown by the blue solid lines of the top, middle, and bottom panels of Fig. 7(b).\\
\hspace*{0.4cm}We compared similar calculations with FEM results for $s/s_0=$ 3 and 4 (not shown) in order to estimate the inter-valley scattering coefficients which were found to be $\left|r\right|^2 =$ 25$\%$, 6$\%$ and 36$\%$ in modes A, B, and C, respectively. The analytical approach indicates that the ring circulation loss in mode B is lower than in modes A and C. The inter-valley scattering could be caused by local perturbations at the corners of the hexagonal ring resonator and the elastic anisotropy of the GaAs device medium. The former is governed by the group velocity, whereas the latter is dependent on the relationship between the wavelength and the side length of the hexagonal ring. Figures 7(c) and 7(d) show the frequency dependence of the group velocity and $\sin{(k L_e/6)}$ (phonon wavenumber $k$) calculated using FEM-simulated dispersion relations, where the black and gray solid lines are valley topological edges of V1/V2 and V2/V1, respectively (see also Fig. 7(a)). The group velocity ($v_g$) in the V1/V2 edge varies moderately with frequency over most of the simulated range, whereas it is drastically reduced beyond 0.510 GHz in the V2/V1 edge. Therefore, the inter-valley scattering arising from the GaAs elastic anisotropy could be large in mode C. In addition, finite backscattering at the corners is a possible origin of standing wave modes (SWM) in such a hexagonal ring resonator. We estimated the frequency condition to sustain SWMs, i.e. $\sin{(k L_e/6)}=0$ in Fig. 7(d); the condition is nearly satisfied with the V2/V1 edge and at the frequencies around mode A and C and thus a relatively large inter-valley scattering would be induced in these modes. We can see a sign of this phenomenon in Fig. 4(a).\\
\begin{figure*}[b]
	\begin{center}
		\vspace{-0.cm}\hspace{-0.0cm}
		\includegraphics[scale=0.6]{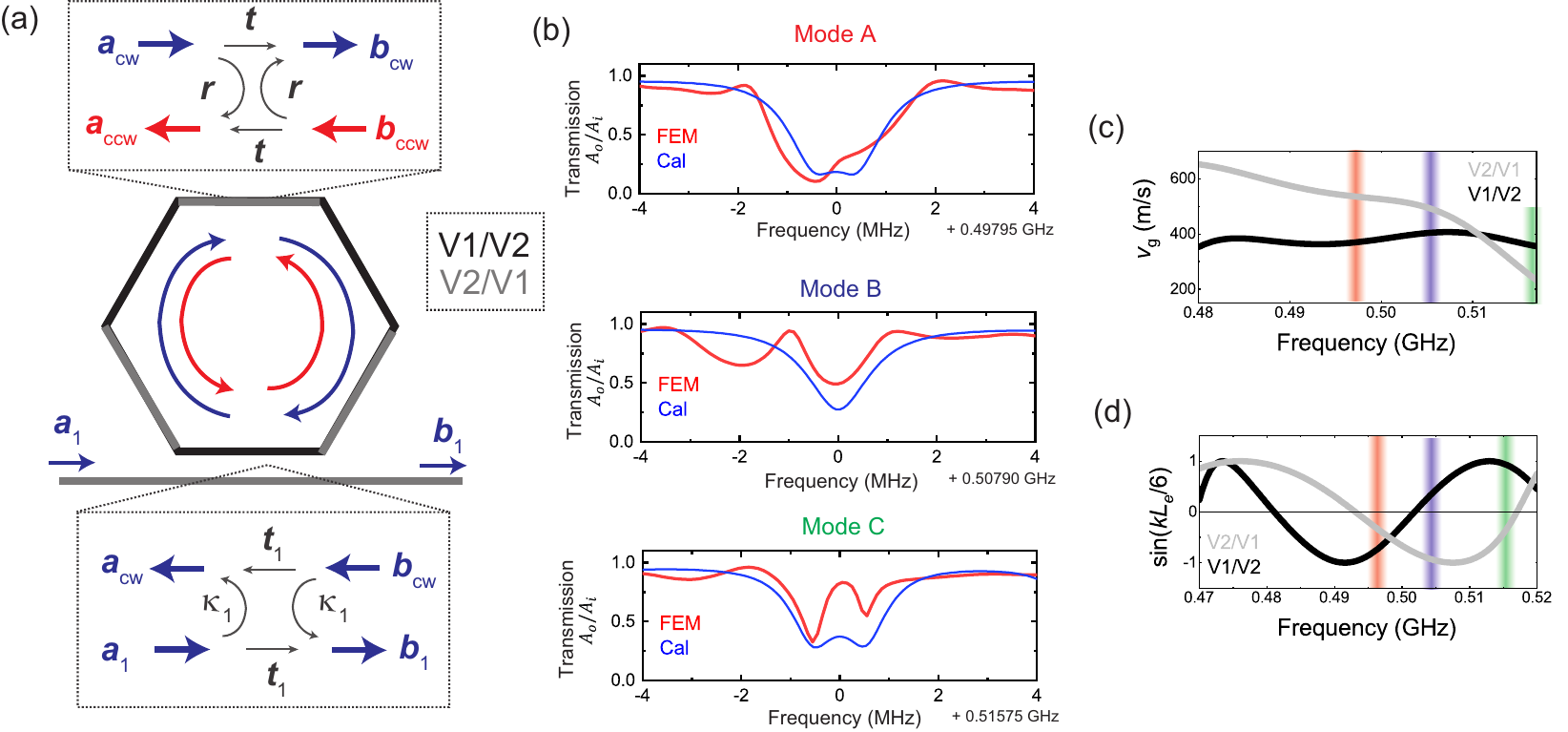}
		\vspace{0.cm}
		\caption{\textbf{(a)} Schematic illustration of analytical simulation model based on the Yariv model. Different colors (black and gray lines) denote different valley edges (V1/V2 and V2/V1). \textbf{(b)} Frequency responses of normalized transmission amplitude in mode A (top), B (middle), and C (bottom) of the VH ring-waveguide coupled systems with $s/s_0=2$, as calculated by FEM (red) and analytically (blue) with (a). Here, the effective ring length and ring-waveguide coupling coefficient are $L_e=$ 54 $\mu$m and $\kappa_1 = e^{-\beta(s-s_0)/s_0}$ with $\beta= 7 \sim 8 \times 10^4$, respectively. $\alpha = \pi f/(Qv_g)$ is the propagation loss, where $Q$ is the intrinsic quality factor of the ring $Q$ = 2,000 and $v_g$ is the group velocity. \textbf{(c)} and \textbf{(d)} Frequency dependence of $v_g$ and $\sin{(k L_e/6)}$ in the V1/V2 (black) and V2/V1 (gray) VH topological edges. The regions highlighted by red, blue, and green are the frequencies of modes A, B, and C.}
		\label{fig S1}
		\vspace{-0cm}
	\end{center}
\end{figure*}
\section{Frequency response of topological hexagonal ring resonator}
\hspace*{0.4cm}A VH hexagonal ring resonator without a coupling to a topological edge waveguide was designed and the spectral characteristics were investigated with FEM, as shown in Fig. 8(a). Removing the waveguide allowed us to ignore possible scattering occurring at the ring-waveguide coupling. There are three ring resonances of modes A, B, and C, as shown in Fig. 8(b). Even with only the ring resonator, distinct dual-peak structures appear in modes A and C. The feature still exists in mode B but is smaller, which could have been caused by scattering arising from the elastic anisotropy of GaAs.\\
\begin{figure*}[t]
	\begin{center}
		\vspace{-0.cm}\hspace{-0.0cm}
		\includegraphics[scale=1.2]{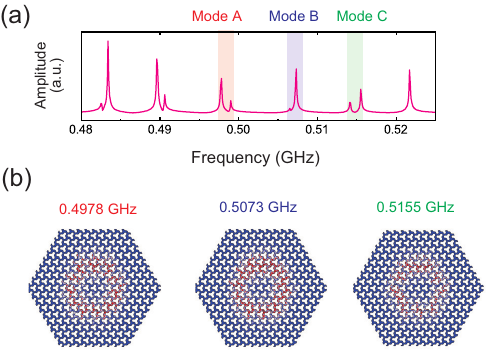}
		\vspace{0.cm}
		\caption{\textbf{(a)} Ring vibration amplitude as function of frequency. \textbf{(b)} Spatial mode profile at 0.4978 GHz (mode A), 0.5073 GHz (mode B), and 0.5155 GHz (mode C).}
		\label{fig S2}
		\vspace{-0cm}
	\end{center}
\end{figure*}
\section{Topological hexagonal ring resonator with isotropic GaAs layer}
\hspace*{0.4cm}We performed FEM simulations on topological ring-waveguide systems based on an isotropic GaAs layer to reveal the effect of the rotation symmetry mismatch between the phononic lattice and the GaAs crystal. The spectral responses in the ring and waveguide are shown in the top and bottom panels of Fig. 9(a). Obviously, both the whispering-galley mode (WGM) and standing-wave mode (SWM) appear in the ring resonant dynamics. In the WGMs, the phase of the elastic waves in the ring continuously varies and the waves circulate in unidirectionally as shown in Fig. 9(b). On the other hand, the SWMs show distinct splitting of the peak spectra with significant inter-valley scattering at the corners.\\
\begin{figure*}[t]
	\begin{center}
		\vspace{-0.cm}\hspace{-0.0cm}
		\includegraphics[scale=1.0]{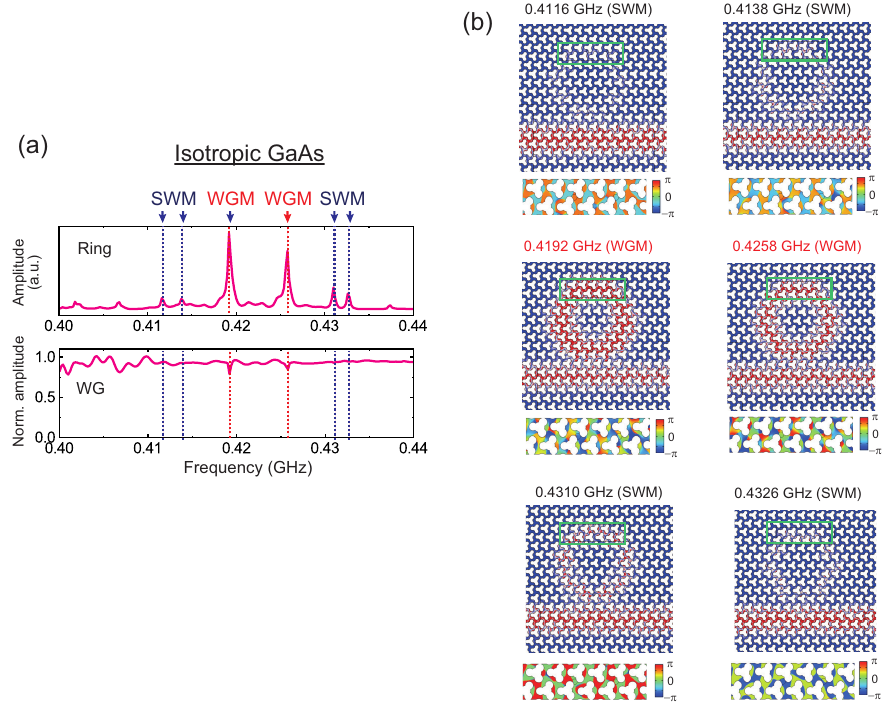}
		\vspace{0.cm}
		\caption{\textbf{(a)} Frequency responses of ring (top) and normalized waveguide transmission (bottom) in an FEM-simulated topologically coupled system with $s/s_0=$3 where the host material is isotropic GaAs with Young's modulus $E=$ 82.68 GPa and Poisson ratio $\nu=$ 0.31. \textbf{(b)} Spatial mode profiles of WGMs and SWMs in the spectral responses, denoted by dotted lines in (a). Phase distributions of these modes in the ring (highlighted in the green box) are shown in the bottom insets.}
		\label{fig S3}
		\vspace{-0cm}
	\end{center}
\end{figure*}
\end{appendix}
\end{document}